\documentclass[preprint,12pt,3p]{elsarticle}
\usepackage{graphicx}
\usepackage{subcaption}
\usepackage{float}
\usepackage{adjustbox}
\usepackage{tabularx,multirow,array}
\usepackage{booktabs}
\usepackage[table]{xcolor}
\usepackage{mwe}
\usepackage{multicol}

    \makeatletter
    \def\ps@pprintTitle{%
       \let\@oddhead\@empty
       \let\@evenhead\@empty
       \let\@oddfoot\@empty
       \let\@evenfoot\@oddfoot
    }
    \makeatother

\usepackage{amssymb}
\usepackage{amsmath}
\usepackage{lineno}

\begin{document}

\begin{frontmatter}

\title{Signature of intermittency in hybrid UrQMD-hydro data at 10 AGeV Au+Au collisions}

\author{Somen Gope}



\ead{somengope30@gmail.com}

\author{Buddhadeb Bhattacharjee\tnoteref{lebel4}}
\tnotetext[lebel4]{Corresponding author.}
\address{Nuclear and Radiation Physics Research Laboratory, Department of Physics, Gauhati University, Guwahati-781014, India}
\ead{buddhadeb.bhattacharjee@cern.ch}


\begin{abstract}
An attempt has been made, in the light of scaled factorial moment (SFM) analysis, to investigate hybrid UrQMD-hydro generated events of Au+Au collisions at 10 AGeV to realize the role of hydrodynamic evolution on observed intermittency, if any. $ln<F_{q}>$ values for $q=2-6$ are found to increase with increasing values of $lnM^{2}$ indicating unambiguously the presence of intermittency in our data sample generated with both chiral and hadronic equation of state (EoS). Although various late processes like meson-meson (MM) and meson-baryon (MB) hadronic re-scattering and/or resonance decays are found to influence the intermittency index significantly, these process would not be held responsible for the observed intermittency in hybrid UrQMD-hydro data.
\end{abstract}


\end{frontmatter}


\section{Introduction}
\label{sec1}
Correlation among the produced particles of relativistic nuclear collisions and fluctuation in number density in phase space bins of such produced particles are believed to be genetically connected. The genesis of the observed correlated emission of produced particles lie on a number of physical processes taking place at various stages of evolution of such nuclear collisions. Different physical processes such as cascading particle production, Bose-Einstein correlation, collective flow, binary decays etc. give correlation among varying number of particles ~\cite{paper1,paper2,paper3,paper4,paper5,paper6,paper7}.\par

Correlated emission of produced particles of a nuclear collision results in preferential emission of particles over some preferred phase space bins of pseudorapidity ($\eta$), azimuthal angle ($\phi$), transverse momentum ($p_{T}$) or any combination of these resulting genuine non-statistical (dynamical) fluctuation in the single particle density distribution spectra. A number of mathematical tools are available to extract and analyze these dynamical fluctuations to have an insight into the collision dynamics in general and particle production mechanism in particular.\par 

Scaled factorial moment (SFM), $F_{q}$, where $q$ is the order of the moments, is one such tested and widely used mathematical tool, discussed in further detail in ref.~\cite{paper4,paper8,paper81}, that filters out dynamical fluctuation from the mixture of statistical and dynamical one. A power law behavior of $F_{q}$ on diminishing phase space bin width $\delta w$, or otherwise, on increasing number $M$ of bins into which the phase space is divided, that is, $F_{q}\propto M^{\alpha_{q}}$ is termed as intermittency where the exponent $\alpha_{q}$ is called the intermittency index and denotes the strength of intermittent particle emission. Intermittency is a property connected with the scale invariance of the physical process and was used first in connection with the turbulence burst in classical hydrodynamics ~\cite{paper30,paper31,paper27}. \par

In the study of the heavy-ion collisions, hydro-dynamical approach remains the heart of the dynamical modeling of such collisions ~\cite{paper32}. Hydrodynamics plays an important role in connecting the static aspects of the matter formed in the collision and the dynamical aspect of such collisions. The high elliptic flow values that have been observed at Relativistic Heavy Ion Collider (RHIC) that seems compatible with some ideal hydrodynamic prediction added additional importance to this approach during last decade ~\cite{paper9}. Hydrodynamics is applied to matter under local equilibrium in the intermediate stage of the evolution of heavy-ion collision (HIC). In this approach a local correspondence between the energy density and the multiplicity of the final hadrons is assumed.\par
The Ultra-Relativistic Quantum Molecular Dynamics (UrQMD) is a QCD based microscopic transport model of nuclear collision that is based on the phase space description of such collisions ~\cite{m_b}. While at low energies collisions are better described in terms of hadronic interactions and resonance decays, at relativistic and ultra-relativistic energies quark and gluons degrees of freedom are introduced via the excitation and fragmentation of strings. The model has found to be highly successful in describing the experimental results of pp, pA and AA collision from SIS to LHC energies ~\cite{paper10,paper11,paper12}. Several attempts ~\cite{paper13,paper141} have been made to investigate non-statistical fluctuation using SFM technique with UrQMD generated data for different systems. However, such analysis of UrQMD data does not exhibit any signature of intermittency ~\cite{paper13,paper14}. \par

UrQMD-hydro, on the other hand, is a hybrid  micro plus macro approach that incorporates transport and hydrodynamical description of heavy-ion collisions for more consistent portrayal of such events from the initial state of collision to final decoupling of hadrons. Here, the microscopic transport calculation for initial condition and freeze-out procedure is implemented with intermediate hydrodynamic calculations ~\cite{paper9,m_b}.\par 

In this work an attempt has been made to analyze UrQMD-hydro generated data using scaled factorial moment technique to realize the presence of intermittency, if any, in the data sample and hence to assess if the hydro plays any role on the observed intermittent type of emission of particles produced in a nuclear collision. Keeping in mind the large acceptance of the upcoming Compressed Baryonic Matter (CBM) experiment of Facility for Antiproton and Ion Research (FAIR), Germany as well as the facts that 10 AGeV is the highest achievable energy for A-A collision at SIS100 of FAIR ~\cite{paper1411} and according to hydrodynamical calculation, the deconfinement phase border is first reached around 10 AGeV ~\cite{paper14111}, the present investiation is carried out with Au+Au generated MC data at 10 AGeV.\par

Further, it has been pointed by a number of workers ~\cite{paper14,paper15} that the correlation among the emitted particles of heavy-ion collisions might washout due to a number of effects such as - effect due to dimensional projection from 3D hyperspace to 1D phase space ($\eta$, $\phi$ or $p_{T}$), contribution from the isotropic decay of the metastable resonances etc. Unlike RHIC of BNL, USA and Large Hadron Collider (LHC) of CERN, Switzerland FAIR-CBM experiment is expected to produce a fireball of high net baryon density. In such scenerio, the final state scattering between the produced particles of the collision influences the observables of such collisions significantly. Therefore, in this work an attempt has also been made to see the effect of final stage hadronic interaction and resonance decays on the observed intermittency by switching on and off the meson-meson (MM), meson-baryon (MB) scattering and resonance decays in UrQMD-hydro model.

\vspace*{1cm}
\section{Results}
\label{sec2}
The analysis was initiated by generating equal numbers ($3.02\times10^{4}$) of UrQMD-hydro (default)~\cite{paper16,paper17,paper18} and UrQMD (default)~\cite{paper19,paper20,paper21} Monte Carlo (MC) events for central (0-5\%) Au+Au collisions at 10 AGeV. To examine the applicability of hybrid UrQMD-hydro model at SIS100 energy, another set of MC events for Au+Au collisions at 8 AGeV is generated and the transverse mass ($m_{T}$) spectra of the generated data is compared with the experimental $m_{T}$ -spectra of E895 experiment (Fig. 1(a)). From the ratio of E895 experiment to our generated spectra, as shown in Fig. 1(b), it could be readily seen that both the results agree well thereby justifying the use of hybrid UrQMD-hydro generated data for the present investigation.

In the study of intermittency in one dimension, a pseudorapidity interval $\Delta\eta$ is divided into $M$ bins of equal width $\delta\eta=\frac{\Delta\eta}{M}$.

If $n_{m}$ be the number of particles in the $m^{th}$ bin, where $m$ can take any value from 1 to $M$ (=10, say), the factorial moment $f_{q}$ of order $q$ is defined as ~\cite{paper4, paper26111}- 

\begin{equation}
\begin{aligned}
f_{q}=<n_{m}(n_{m}-1).......(n_{m}-q+1)>
\end{aligned}
\end{equation}

If the averaging in the above equation is performed over all events for a fixed bin, the procedure is called vertical averaging and gives fluctuation in event space. On the other hand, if $n_{m}$ is averaged over all bins for a fixed event, it is called horizontal averaging and provides information on fluctuation in phase space.

\begin{figure}[H]
\begin{subfigure}{0.5\textwidth}
  \centering
  \includegraphics[width=0.95\linewidth]{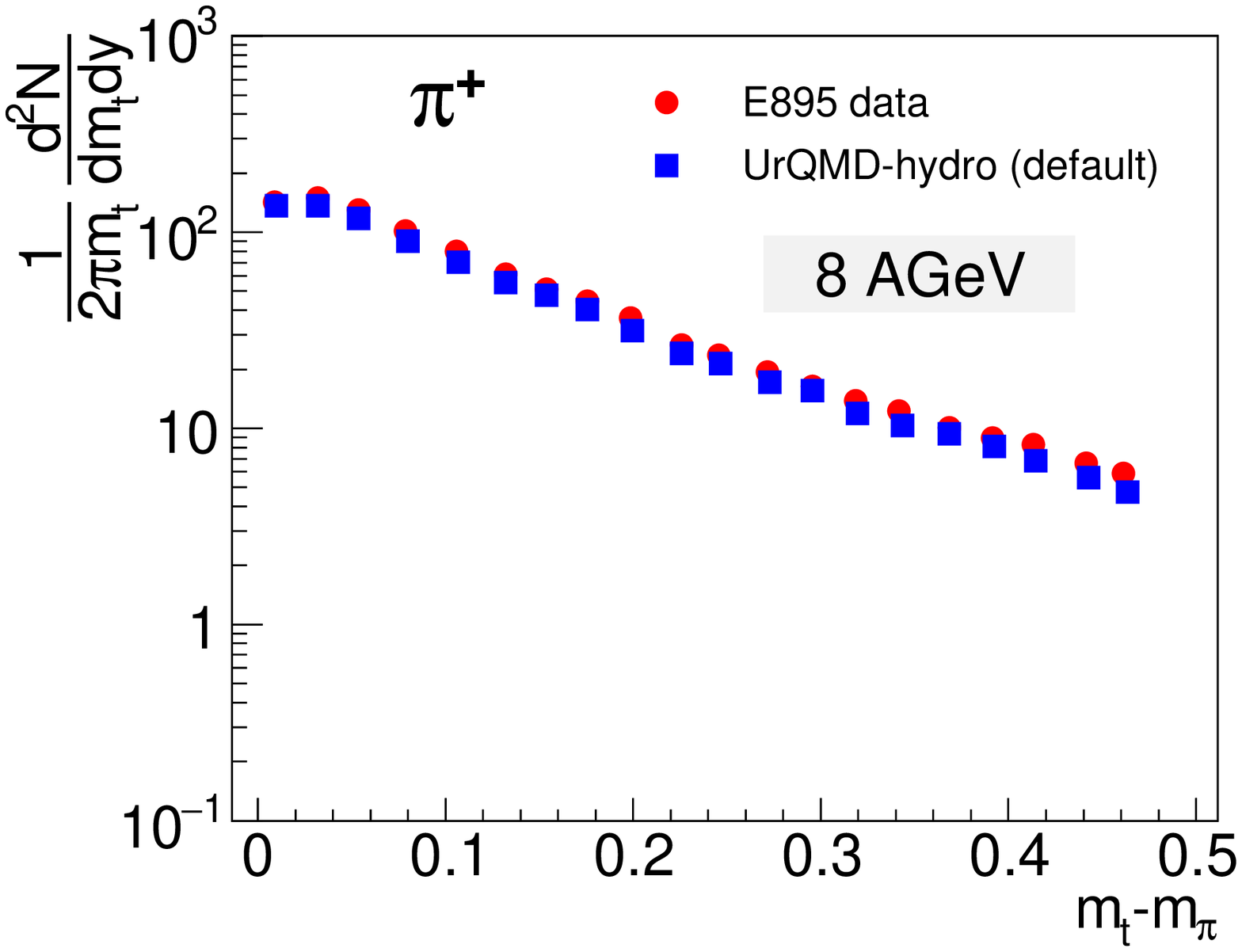}
  \caption{}
  \label{fig1:sub1}
\end{subfigure}%
\begin{subfigure}{0.5\textwidth}
  \centering
  \includegraphics[width=1.05\linewidth]{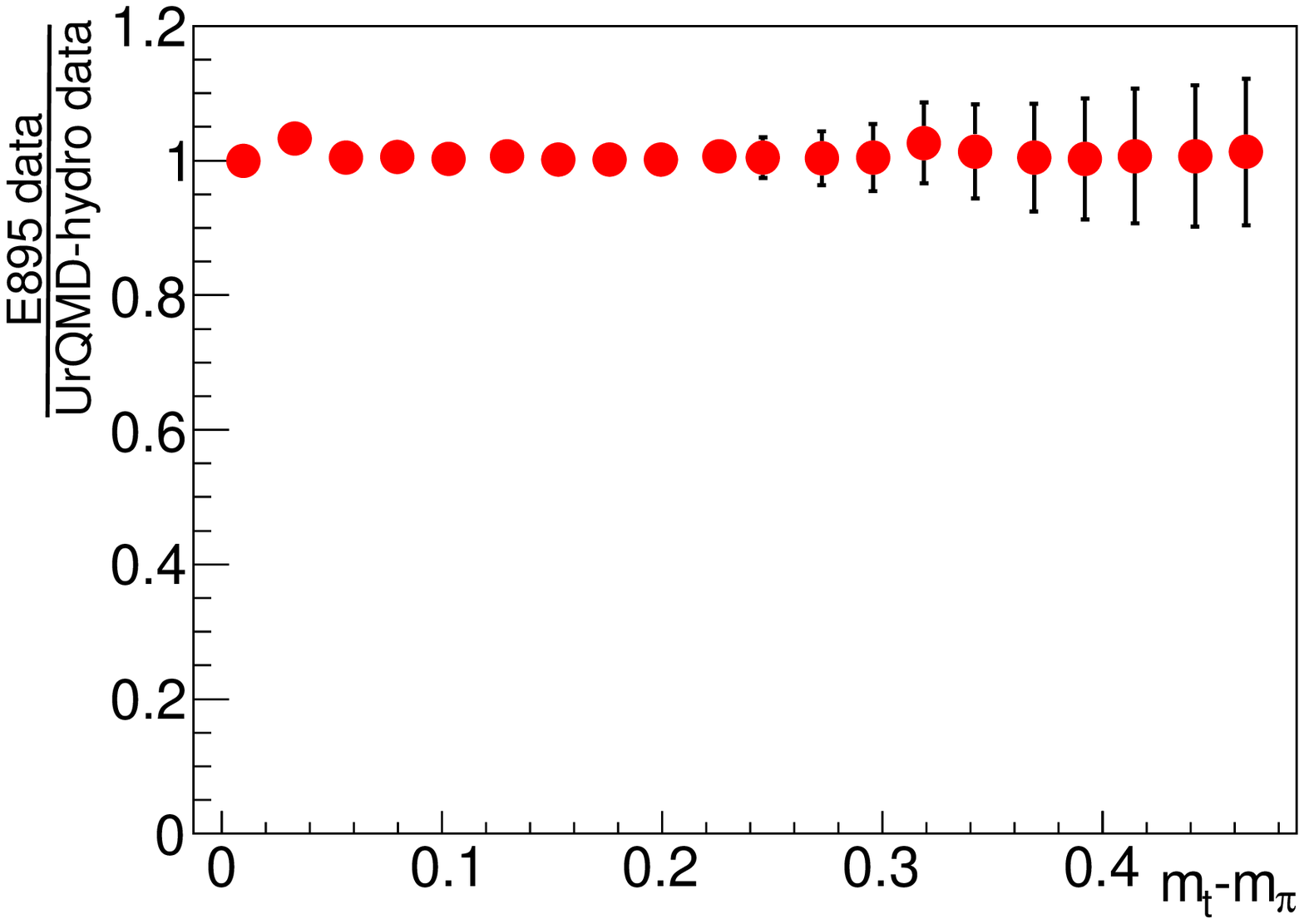}
  \caption{}
  \label{fig2:sub2}
\end{subfigure}
\caption{Transverse mass spectra of $\pi^{+}$ with UrQMD-hydro data for Au+Au collisions at 8 AGeV (a) compared with experimental data of E895 ~\cite{paper211}  and (b) ratio of $m_{T}$-spectra of $\pi^{+}$ between experimental data to model generated MC data.}
\end{figure}

Assuming that the statistical contribution to the fluctuation in the spatial distribution of the charged particles is Poisson distributed, Bialas and Peschanski ~\cite{paper4, paper8} showed that the factorial moments of the multiplicity distribution of the entire sample of events are equivalent to the moments of the corresponding dynamical part only, irrespective of the nature of the statistical component. In either method of averaging, if the probability distribution $P_{n}$ of $n_{m}$ can be expressed as a convolution of dynamical distribution $D(\nu)$ and the statistical (Poissonian) distribution, $f_{q}$ is shown to be a simple moment of $D(\nu)$, the statistical component is regarded as having been filtered out by $f_{q}$ estimation ~\cite{paper2111}.

For a single event, the $q^{th}$ order scaled factorial moment is defined as-

\begin{equation}
\begin{aligned}
F_{q}=M^{q-1}\sum_{m=1}^{M}\frac{n_{m}(n_{m}-1)...(n_{m}-q+1)}{n(n-1)...(n-q+1)}
\end{aligned}
\end{equation}

where, $n$ is the multiplicity of an event. Thus, $n=\sum_{m=1}^{M} n_{m}$ and $\frac{\chi_{max}-\chi_{min}}{M} = \frac{1}{M}$. 

For an ensemble of events having varying multiplicity, the expression for scaled factorial moment is modified as-

\begin{equation}
\begin{aligned}
F_{q}=M^{q-1}\sum_{m=1}^{M}\frac{n_{m}(n_{m}-1)...(n_{m}-q+1)}{<n>^{q}}
\end{aligned}
\end{equation}

where, $<n>= \frac{\sum_{N_{ev}=1}^{N_{ev}}n}{N_{ev}}$, $N_{ev}$ is the total number of events of the population.

The horizontally averaged normalized or scaled factorial moment is then expressed as-

\begin{equation}
\begin{aligned}
<F_{q}>=\frac{1}{N_{ev}}\sum_{i=1}^{N_{ev}}M^{q-1}\sum_{m=1}^{M}\frac{n_{m}(n_{m}-1)...(n_{m}-q+1)}{<n>^{q}}
\end{aligned}
\end{equation}

To minimise the projection effect, if any, the analyses of the UrQMD-hydro and UrQMD data using SFM technique were carried out in two dimensional pseudorapidity-azimuthal space. Initial shape dependence of the two dimensional density distribution spectrum (Fig. 2(a) and (b)) is removed by converting the pseudorapidity ($\eta$) and azimuthal angle ($\phi$) values of every primary charged particle of each generated event to a new cumulative variable $\chi(\eta)$ and $\chi(\phi)$ respectively, defined as -

\begin{equation}
\begin{aligned}
\chi(\eta) = \frac{\int_{\eta_{min}}^{\eta}\rho(\eta)d\eta}{\int_{\eta_{min}}^{\eta_{max}}\rho(\eta)d\eta}   \qquad  \textrm{ } \textrm{and} \textrm{ }    \qquad  \chi(\phi) = \frac{\int_{\phi_{min}}^{\phi}\rho(\phi)d\phi}{\int_{\phi_{min}}^{\phi_{max}}\rho(\phi)d\phi}
\end{aligned}
\end{equation}
where, $\eta_{min}$ = -5.0, $\eta_{max}$ = 5.0 , $\phi_{min}$ = 0 and $\phi_{max}$ = 6.28. $\chi(\eta)$ and $\chi(\phi)$ vary from 0 to 1. The two dimensional $\chi(\eta-\phi)$ space is now divided into $M_{i}$ $\times$ $M_{i}$ bins of equal width $d\chi_{\eta}$ $\times$ $d\chi_{\phi}$ where $M_{i}$=1 to 10 and $d\chi_{\eta}=\frac{\chi_{max}(\eta)-\chi_{min}(\eta)}{M}=\frac{1}{M}$ and $d\chi_{\phi}=\frac{\chi_{max}(\phi)-\chi_{min}(\phi)}{M}=\frac{1}{M}$ ~\cite{paper261, paper2611}. Obviously, the minimum and maximum values of $d\chi_{\eta}$ (and $d\chi_{\phi}$) would be 0.1 and 1 respectively. Thus, the size of the smallest bin of the two dimensional $\chi(\eta-\phi)$ space would be 0.1 $\times$ 0.1 when it is divided into hundred square bins of equal size. The single particle density distribution spectrum of two dimensional $\eta$-$\phi$ space (Fig. 2(b)) is then replotted in two dimensional $\chi(\eta-\phi)$ cumulant phase space as shown in Fig. 2(c) with UrQMD-hydro (default) and UrQMD (not shown) generated data. It could be readily seen from Fig. 2(c) that, as expected, the distribution is perfectly flat in $\chi(\eta-\phi)$ space and is free from any preferential emission thereby minimizing the scope of any error in our fluctuation studies due to initial (kinematic) shape dependence of the single particle spectra itself. 

\begin{figure}[H]
\begin{subfigure}{0.3\textwidth}
  \centering
  \includegraphics[width=.9\linewidth]{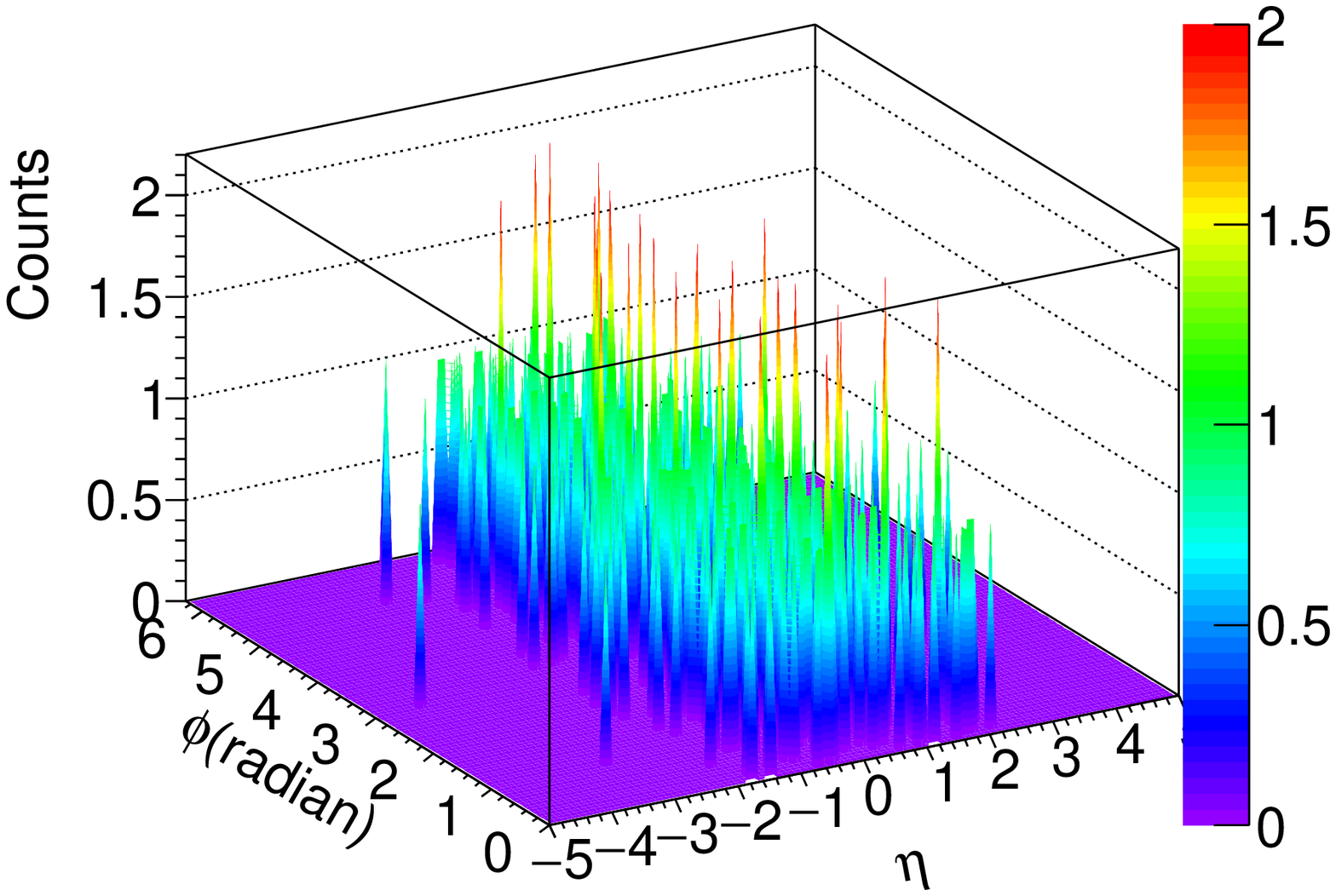}
  \caption{}
  \label{fig3:sub3}
\end{subfigure}%
\begin{subfigure}{0.3\textwidth}
  \centering
  \includegraphics[width=1.0\linewidth]{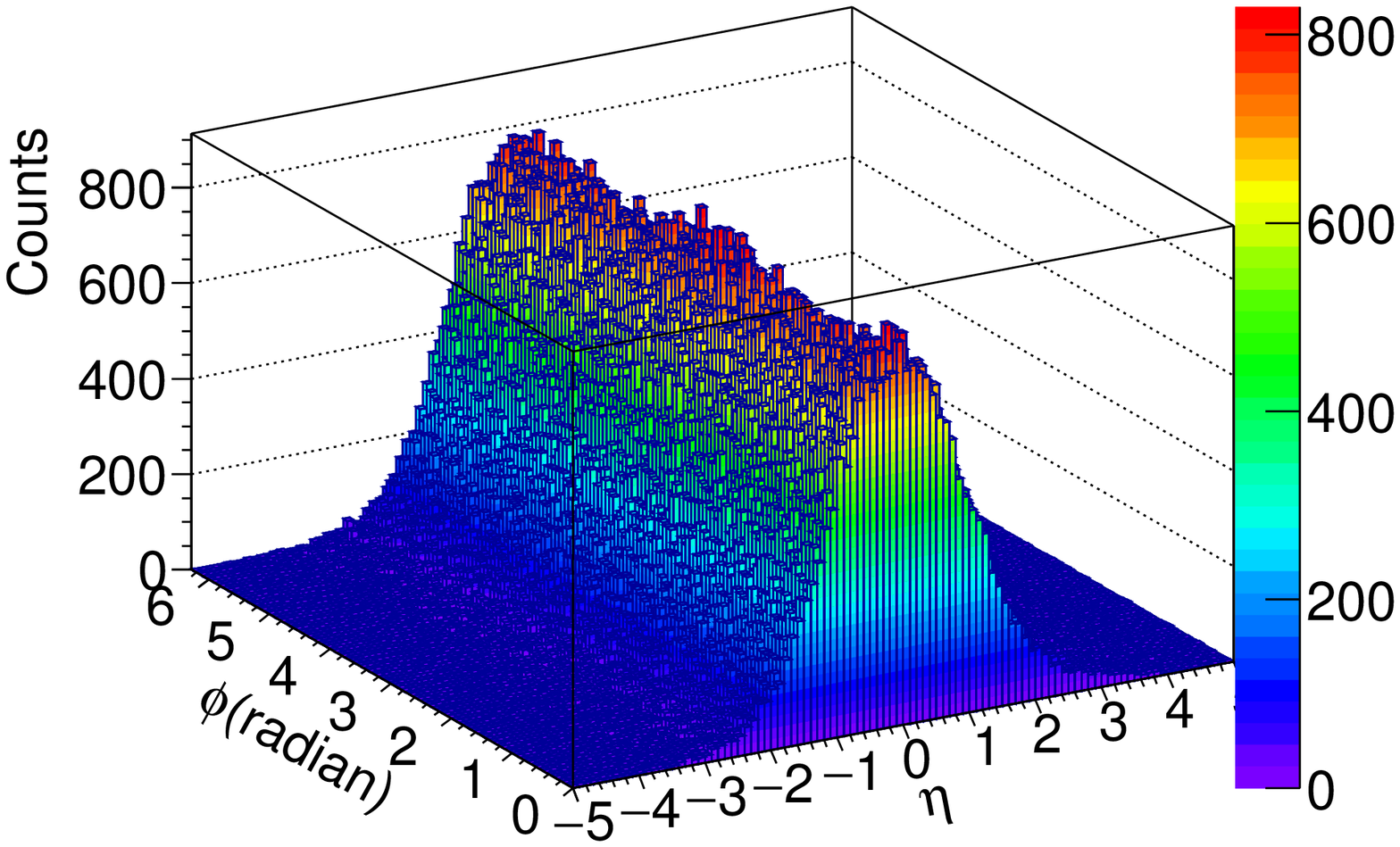}
  \caption{}
  \label{fig4:sub4}
\end{subfigure}
\begin{subfigure}{0.32\textwidth}
  \centering
  \includegraphics[width=1.0\linewidth]{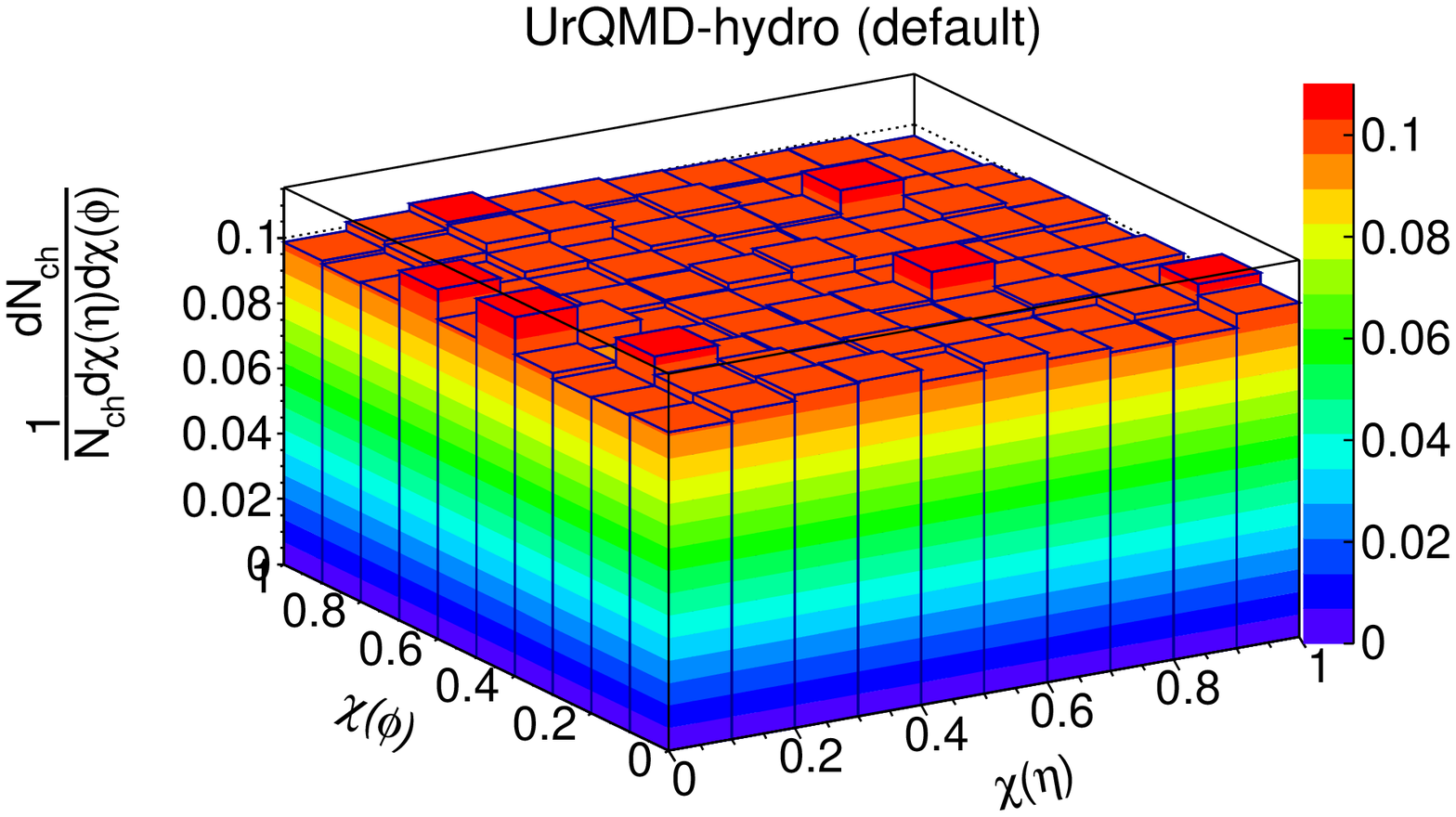}
  \caption{}
  \label{fig5:sub5}
\end{subfigure}
\caption{Density distribution spectrum for (a) a single event in 2D $\eta-\phi$ space, (b) entire sample in 2D $\eta-\phi$ space and (c) entire sample in 2D $\chi(\eta-\phi)$ spaces.}
\end{figure}

Equal number of events are then generated using random number generator (RAN-GEN) with same multiplicity as that of each event of UrQMD-hydro data with $\chi(\eta)$ and $\chi(\phi)$ values for each particle randomly generated between 0 and 1.

To estimate the scaled factorial moment in two dimensional cumulant $\chi(\eta-\phi)$ space using the above formula (Eq. 4), the two dimensional $\chi(\eta-\phi)$ space is successively divided into $M_{i}$ $\times$ $M_{i}$ = $M^{2}$=1, 4, 9, 16, ...., 100 bins of equal width $d\chi_{\eta}$ $\times$ $d\chi_{\phi}$. The number of particles populating each square bin is computed to estimate the corresponding SFM. The SFM estimated for each bin are then averaged for all bins and finally over all events to get $<F_{q}>$ for different values of $M^{2}$.

The two dimensional horizontally averaged scaled factorial moments $<F_{q}>$ of order $q=2-6$ are then estimated for $\chi(\eta-\phi)$ space with UrQMD-hydro, UrQMD and RAN-GEN generated data and $ln<F_{q}>$ is plotted against $lnM^{2}$ in Fig 3(a). From this plot, a clear signature of power law behavior of the form $<F_{q}>\propto M^{\alpha_{q}}$ for the estimated values of $<F_{q}>$ with the increasing number of phase space bin $M^{2}$ could be observed from $ln<F_{q}>$ vs. $ln M^{2}$ plot for $q=2-6$ confirming the presence of intermittency in UrQMD-hydro generated data with chiral EoS.

\begin{figure}[H]
\begin{subfigure}{0.5\textwidth}
  \centering
  \includegraphics[width=1.0\linewidth]{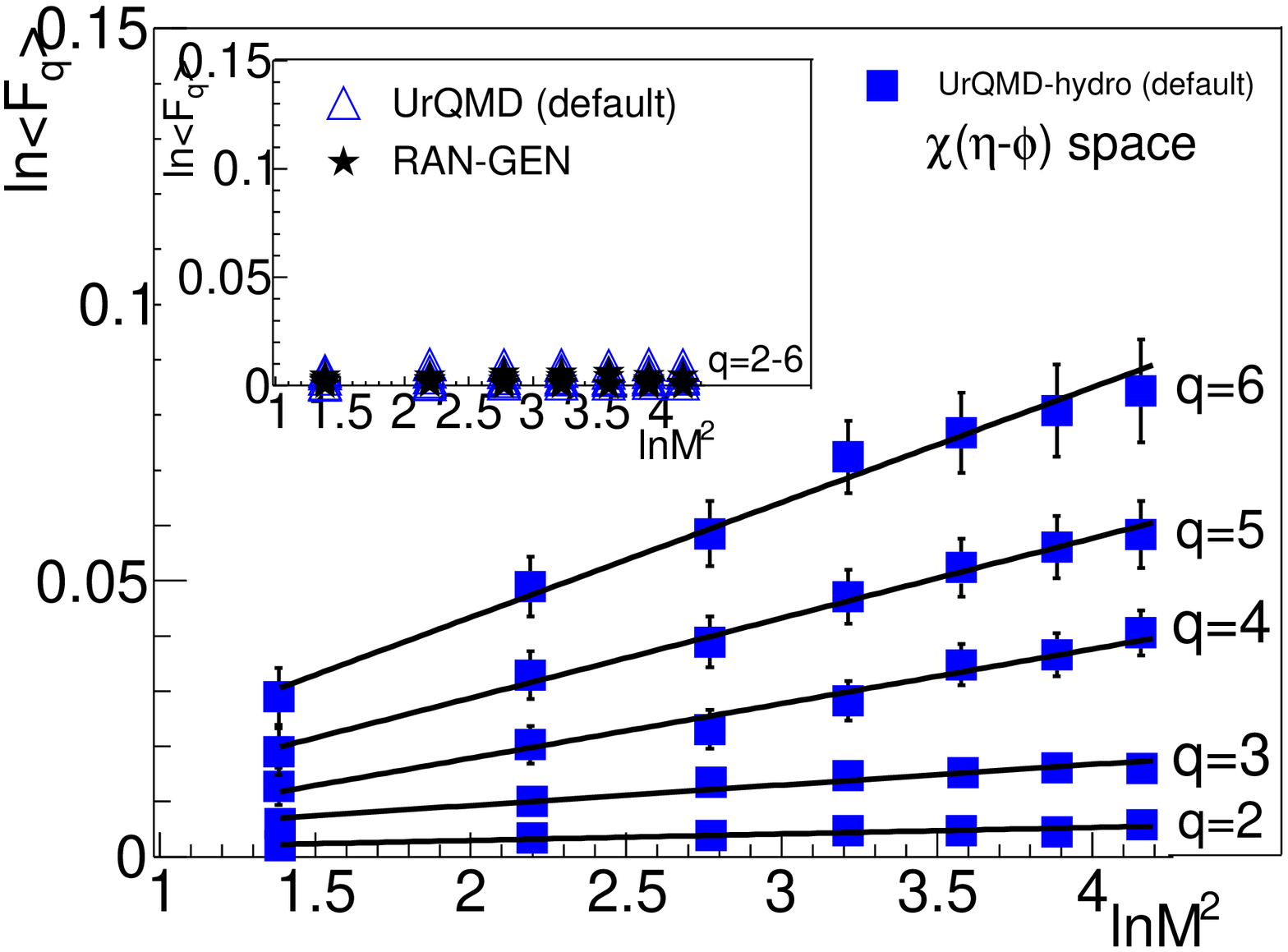}
  \caption{}
  \label{fig6:sub6}
\end{subfigure}%
\begin{subfigure}{0.5\textwidth}
  \centering
  \includegraphics[width=0.9\linewidth]{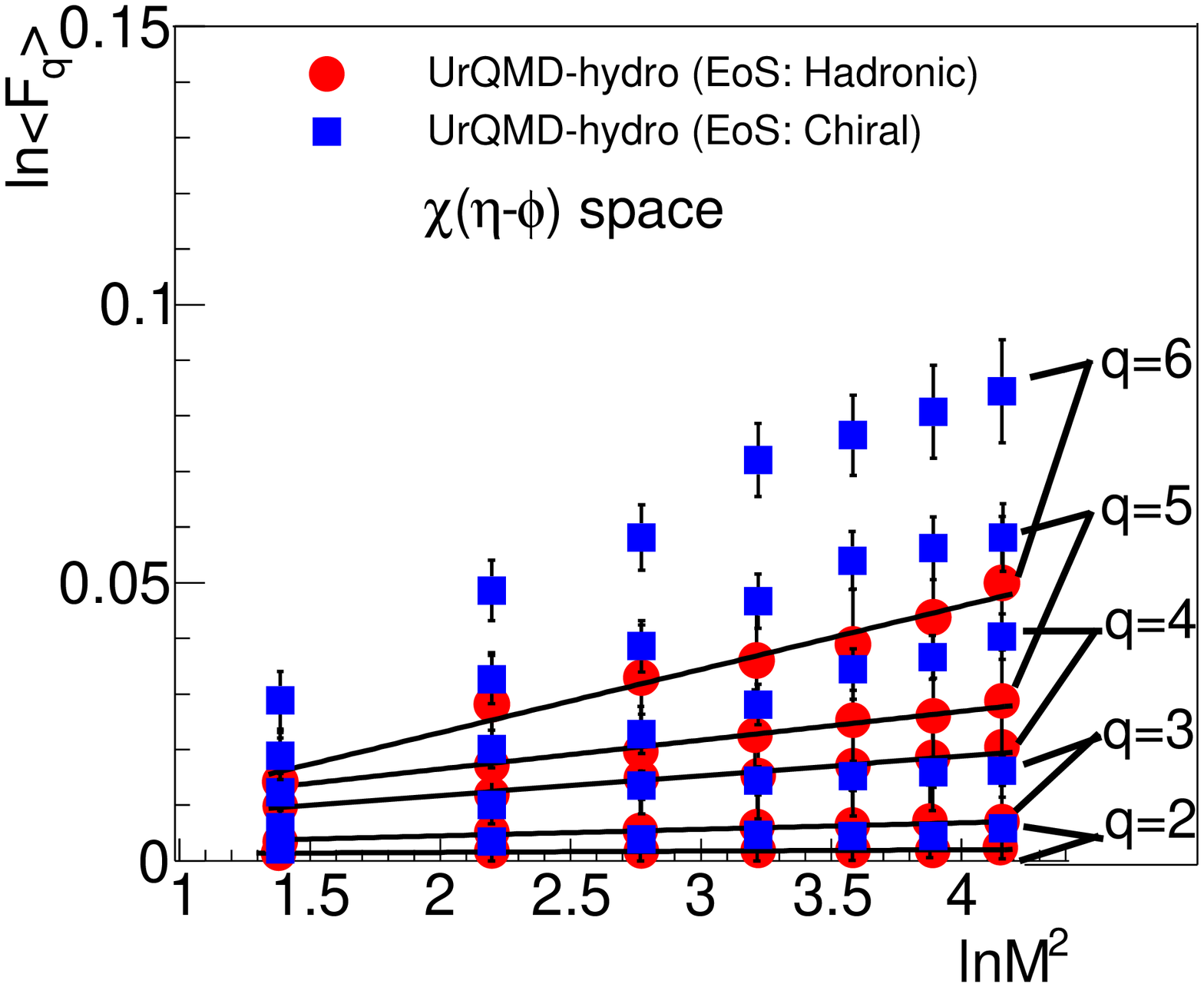}
  \caption{}
  \label{fig7:sub7}
\end{subfigure}
\caption{$ln<F_{q}>$ vs $lnM^{2}$ plots for (a) UrQMD-hydro (default) events and UrQMD and RAN-GEN events (inset) (b) for UrQMD-hydro events with hadronic and chiral Equation of State (EoS). Solid straight lines are the best fitted lines to the data points.}
\end{figure}

However, as evident from the inset plot of Fig. 3(a), no such intermittency effect could be seen with UrQMD (transport model) and RAN-GEN generated data. The observation with our UrQMD set of generated data is consistent with the results reported by other workers ~\cite{paper13,paper141}. Since in UrQMD model no critical function is introduced, SFM analysis does not exhibit any signature of intermittency, whereas intermittency in UrQMD-hydro (default) data could be due hydrodynamic evolution of the matter created in the collisions or/and due to use of chiral EoS.\par

\begin{figure}[H]
\begin{subfigure}{0.5\textwidth}
  \centering
  \includegraphics[width=0.86\linewidth]{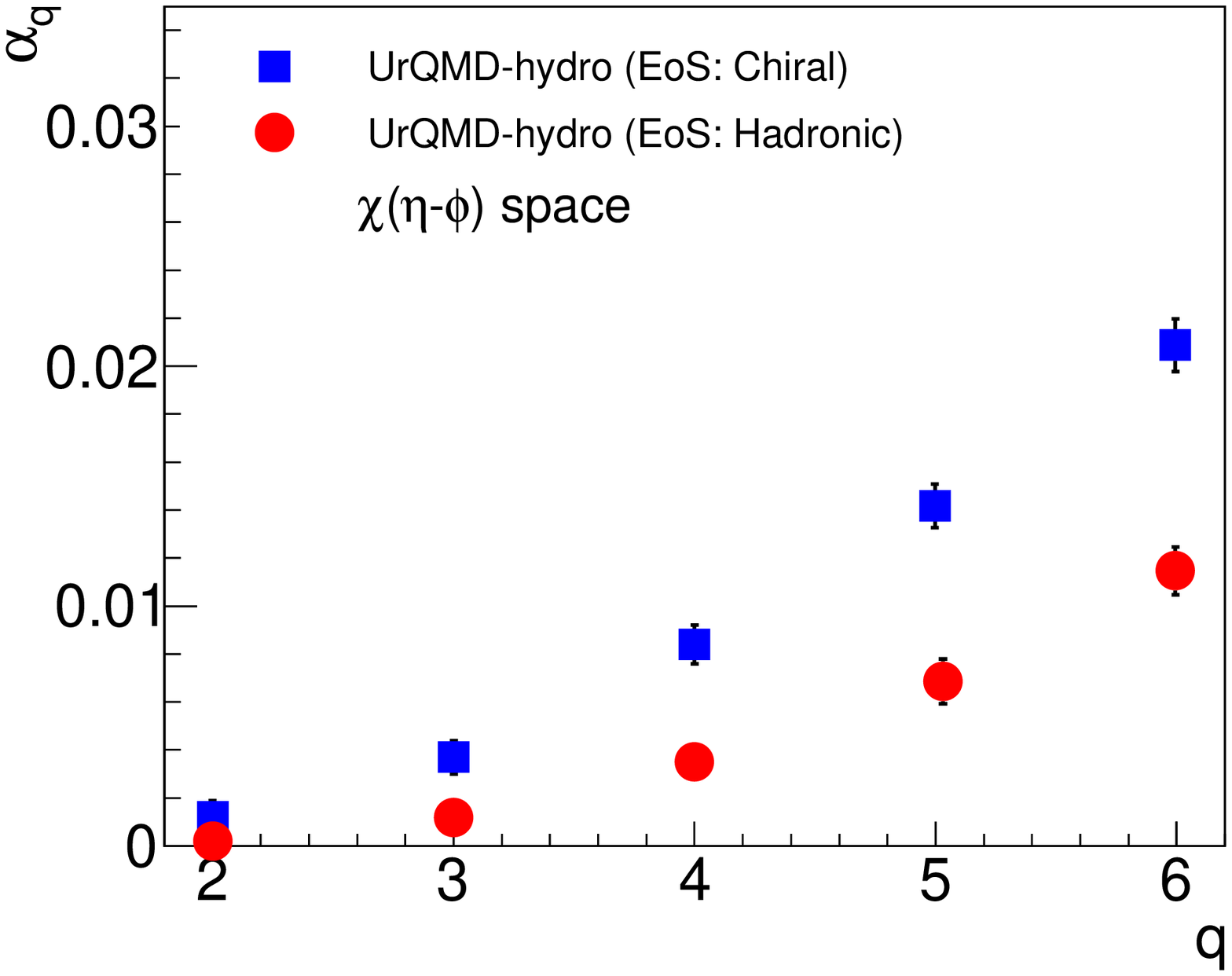}
  \caption{}
  \label{fig8:sub8}
\end{subfigure}%
\begin{subfigure}{0.5\textwidth}
  \centering
  \includegraphics[width=0.98\linewidth]{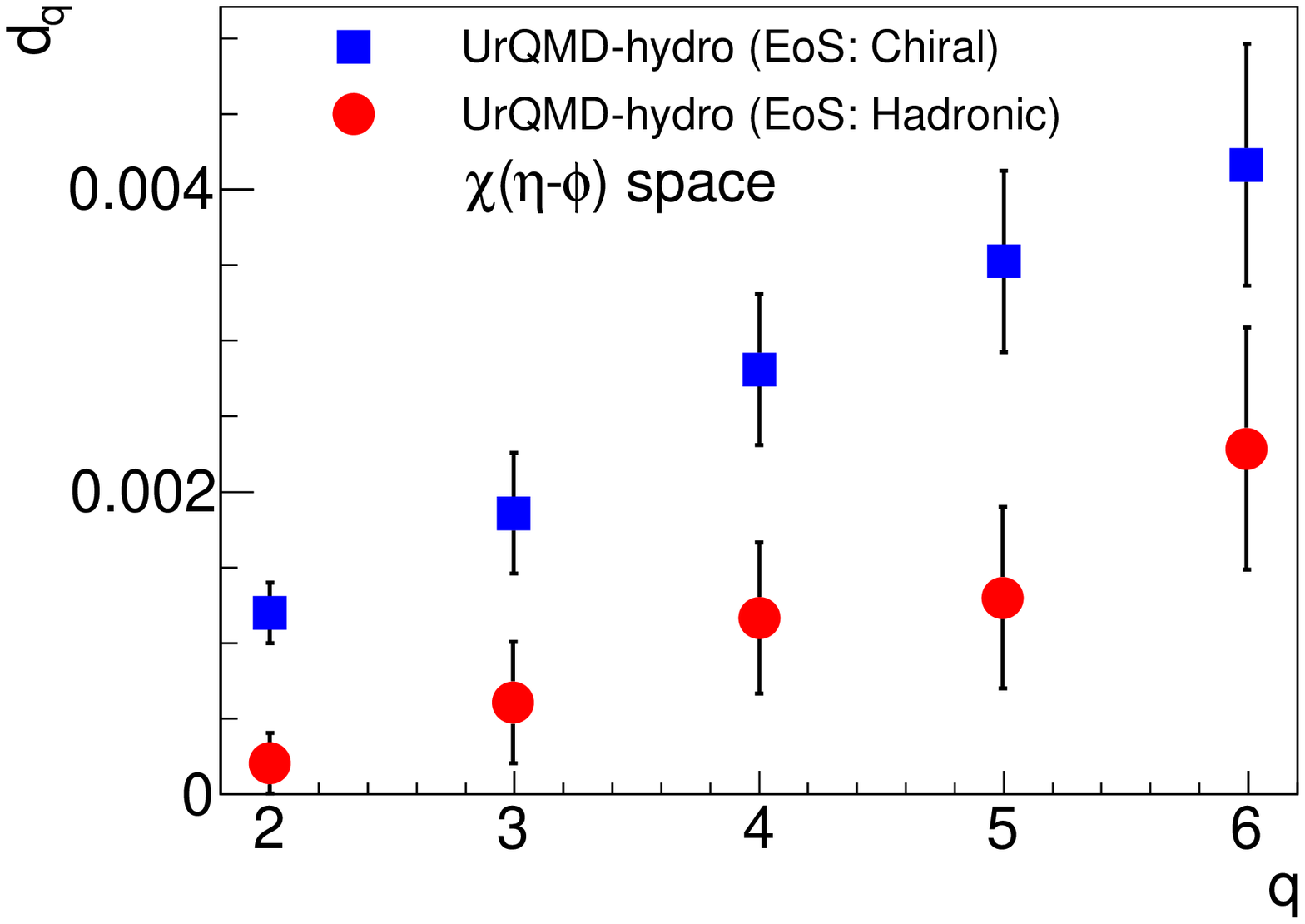}
  \caption{}
  \label{fig9:sub9}
\end{subfigure}
\caption{(a) Intermittency index ($\alpha_{q}$) vs $q$ and (b) Anamalous fractal dimension ($d_{q}$) vs $q$ for UrQMD-hydro data with chiral and hadronic EoS.}
\end{figure}

To ascertain if the observed intermittency in our hydro generated data is due to chiral phase-transition or not, realised in the model through the use of chiral EoS, another set of UrQMD-hydro data was generated with hadronic EoS. The result of 2D analysis is presented in Fig. 3(b) for both the sets of data generated with hadronic and chiral (default) EoS. A clear increase in the values of $ln<F_{q}>$ against $lnM^{2}$ could be seen with UrQMD-hydro central (0-5\%) data for both hadronic and chiral EoS. With chiral EoS, the intermittency indices $\alpha_{q}$ for $q=2-6$ are found to be significantly larger than that of hadronic EoS data. The values of intermittency indices for different order of moments as estimated from this analysis with different sets of data are listed in table 1.  The variation $\alpha_{q}$ against $q$ for UrQMD-hydro generated data with chiral and hadronic EoS are shown in Fig. 4(a). In order to estimate the error of intermittency index $\alpha_{q}$, we have adopted the method of simulating several independent event samples, each of the same size and estimating $\alpha_{q}$ for each sample and also estimating $\alpha_{q}$ with different binning for each sample and then adding the errors of $\alpha_{q}$ due to different sample and different bin width in quadrature ~\cite{paper261}. The observed stronger intermittency in data sample of UrQMD-hydro with chiral EoS than that of hadronic EoS data may be attributed to cascading particle production in partonic media produced due to the use of chiral EoS.\par

Intermittency, in turn, is related to self similarity and fractal behavior of the emission spectra ~\cite{paper4,paper24,paper25,paper271}. The anomalous fractal dimension $d_{q}$ ($= D - D_{q}$ , where $D$ and $D_{q}$ are ordinary topological dimension and generalized fractal dimension respectively),  is related to intermittency index $\alpha_{q}$ through the relation

\begin{equation}
\begin{aligned}
d_{q} = \frac{\alpha_{q}}{(q-1)}
\end{aligned}
\end{equation}

A study on the order $q$ dependence of $d_{q}$ is quite informative about the particle production mechanism. It is claimed that an increase in $d_{q}$ with $q$ is associated with particle production via some branching mechanism. An order independence of $d_{q}$, on the other hand, is indicative of  particle production via a phase-transition. In Fig. 4(b), the variation of $d_{q}$ with $q$ is shown for UrQMD-hydro generated data with both hadronic and chiral equation of states and is found to increase monotonically with the increase of $q$ for both the sets of data. However, $d_{q}$ is consistently found to be larger in data sample with chiral EoS than that of hadronic EoS indicating the fact that particles of UrQMD-hydro data with chiral EoS occupy less phase space than that of hadronic EoS, or otherwise, particle emission is more preferential in partonic media than that of hadronic media. 

\begin{figure}[H]
\begin{subfigure}{0.5\textwidth}
  \centering
  \centerline{\includegraphics[width=0.98\linewidth]{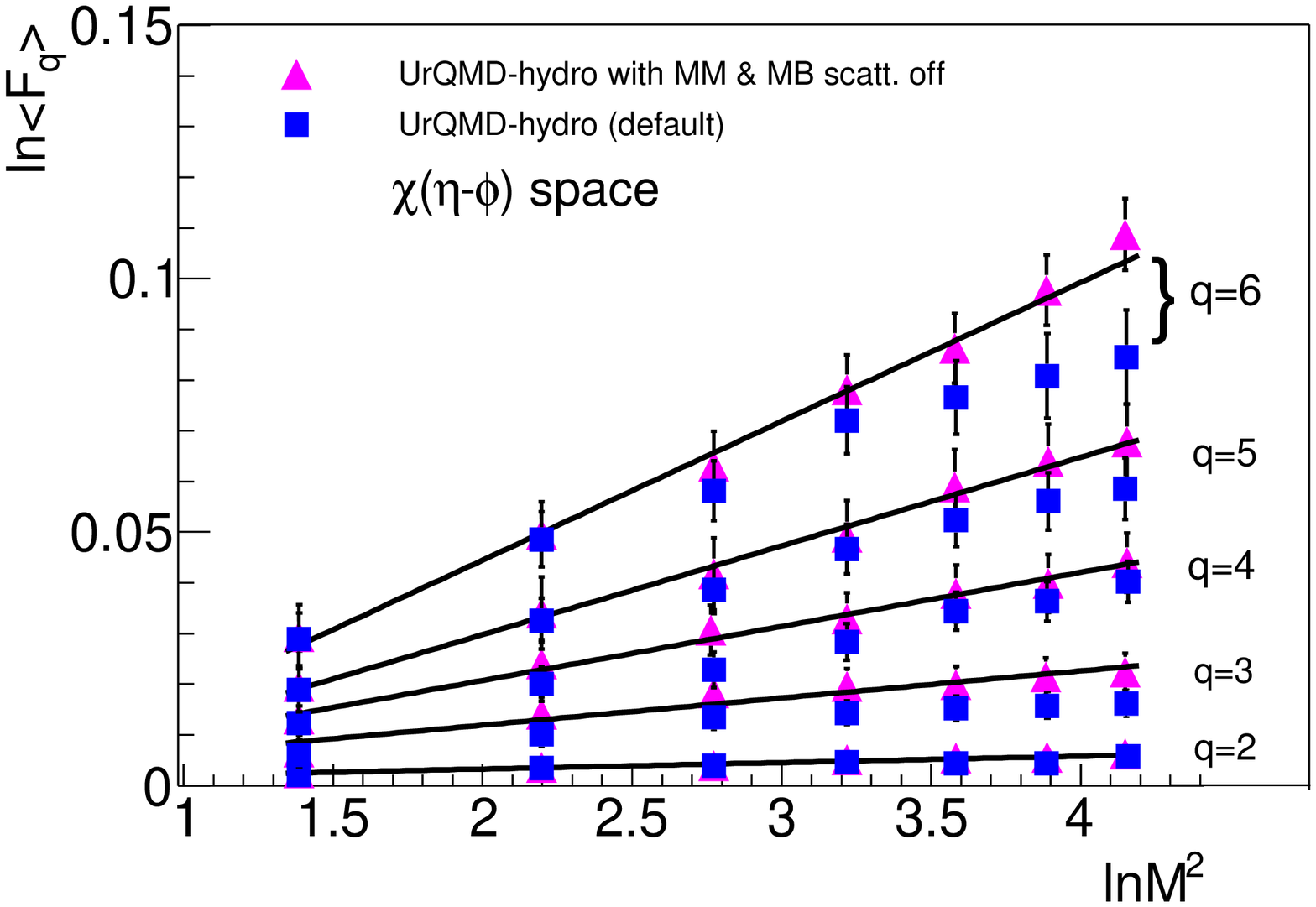}}
  \caption{}
  \label{fig10:sub10}
\end{subfigure}%
\begin{subfigure}{0.5\textwidth}
  \centering
  \includegraphics[width=0.84\linewidth]{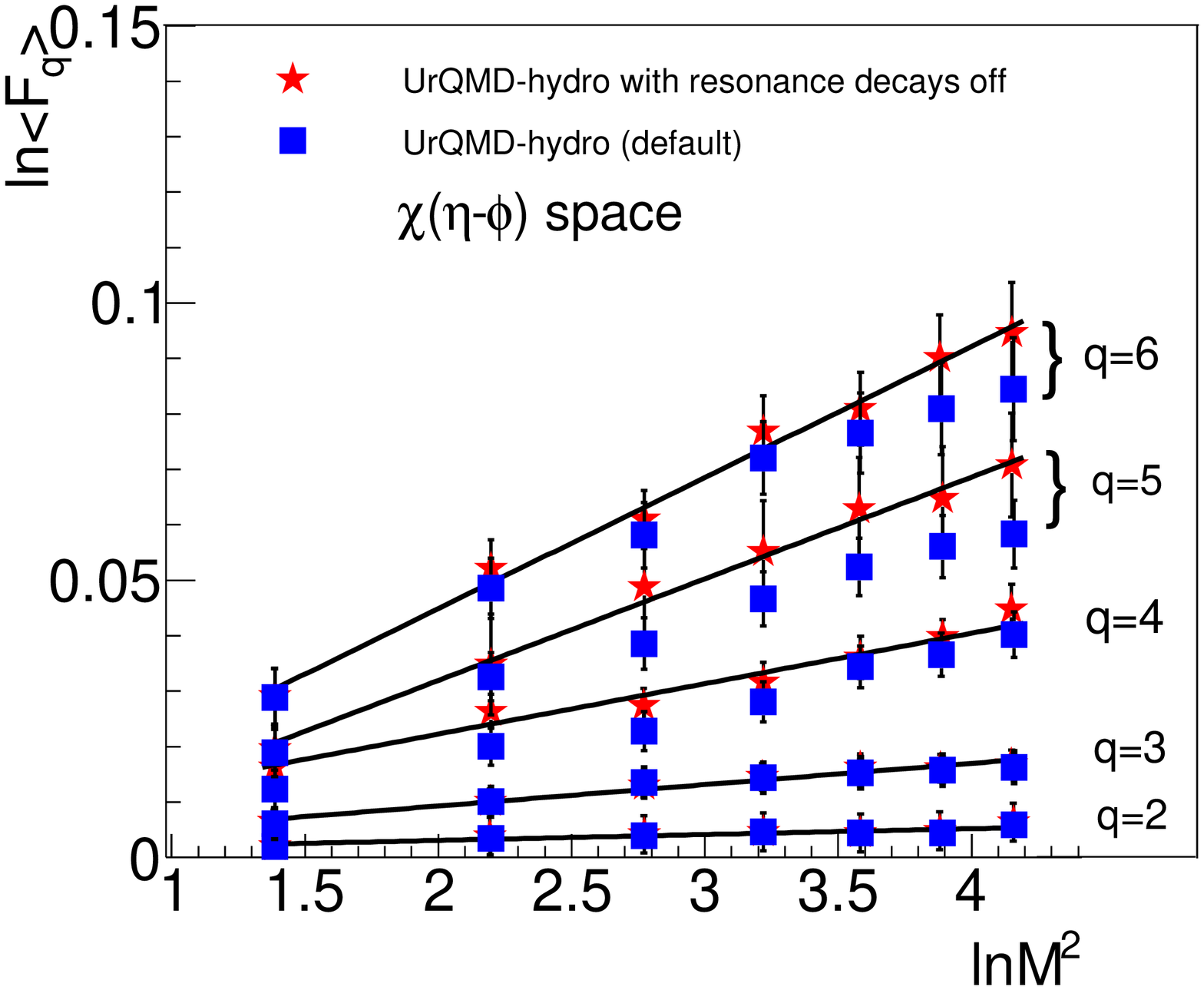}
  \caption{}
  \label{fig11:sub11}
\end{subfigure}
\begin{subfigure}{0.5\textwidth}
  \centering
  \includegraphics[width=0.94\linewidth]{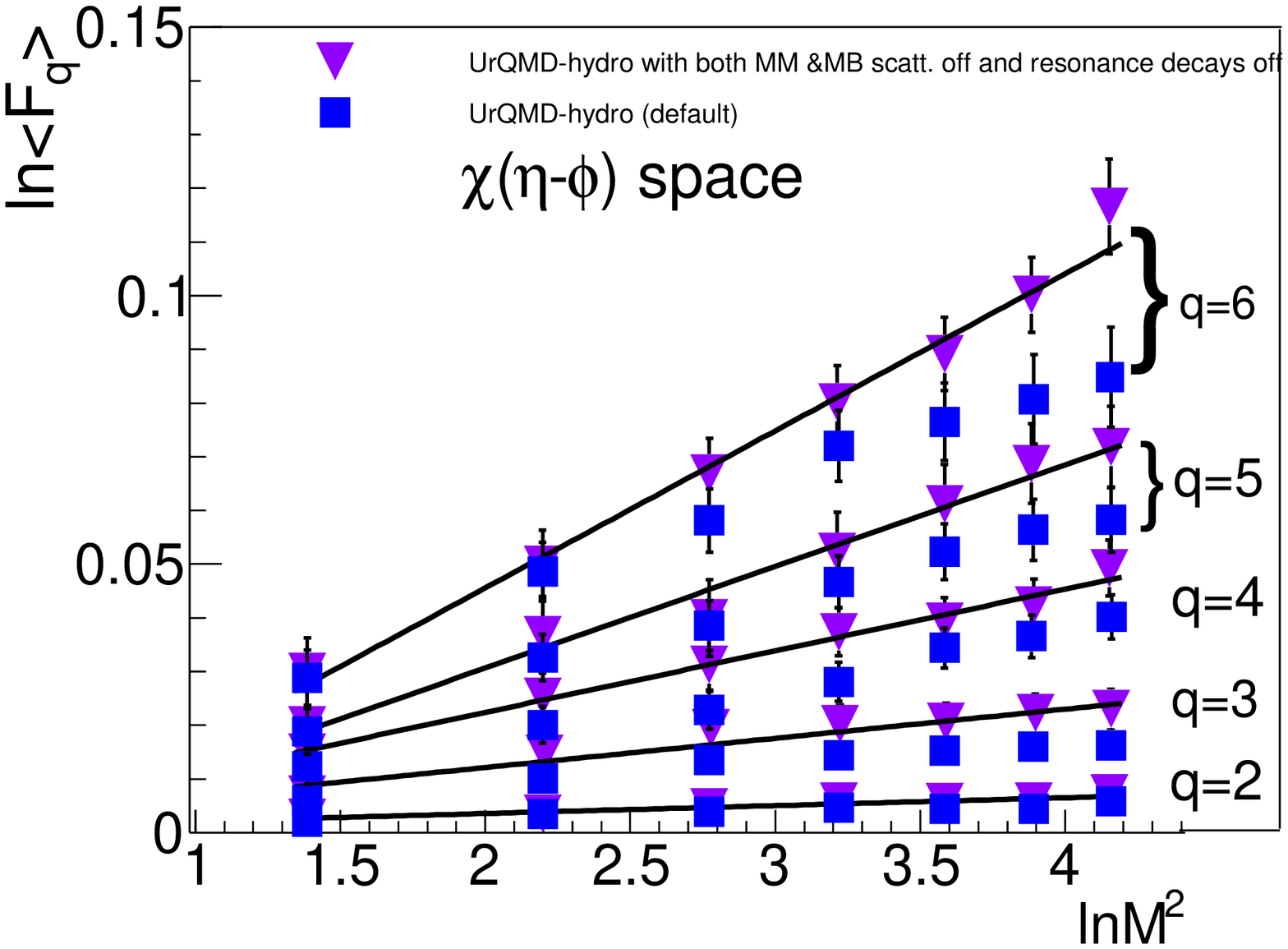}
  \caption{}
  \label{fig12:sub12}
\end{subfigure}
\caption{$ln<F_{q}>$ vs $lnM^{2}$ in $\chi(\eta-\phi)$ space for UrQMD-hydro data with (a) MM and MB scattering off (b) resonance decays off (c) both MM, MB scattering and resonance decays off. Solid straight lines are the best fitted lines for the data points.}%
\end{figure}

The hadronic re-scattering and/or resonance decays have substantial impact on most hadronic observables, such as correlations and fluctuations ~\cite{paper301, paper302}. Experimentally, one measures only final abundances of hadrons which includes both primordial particle production as well as contribution from the resonance decays. Production of resonances plays an important role for studying various properties of interaction dynamics in heavy-ion collisions. Resonances, having short life time that subsequently decay into stable hadrons, as well as hadronic re-scattering can effect the final hadron yeilds and their number fluctuations ~\cite{paper302}. To evaluate the contribution of such processes on the observed intermittency, three new sets of UrQMD-hydro (default) events are generated with $(i)$ meson-meson (MM) and meson-baryon (MB) scattering off but resonance decays on, $(ii)$ MM, MB scattering on but resonance decays off,  and $(iii)$ MM, MB, and resonance decays all off.  $ln<F_{q}>$ vs $lnM^{2}$ plots for all such events are shown in Fig. 5(a), (b) and (c). It is readily evident from these figures that all these late stage processes such as hadronic re-scattering and/or resonance decays weaken the signatures of intermittency considerably. Thus, none of these processes are the cause of observed intermittent type of particle emission in our hybrid UrQMD-hydro generated data.

\begin{table}[]
\caption{Intermittency index values for q=2-6 for various systems using UrQMD-hydro model.}
\centering
\begin{tabular}{p{2.0cm}| l l l l l}
\toprule
Systems & \multicolumn{5}{c@{}}{Intermittency index($\alpha_{q}$)}\\
\cmidrule(l){2-6}
& $\alpha_{2}\times10^{-3}$ & $\alpha_{3}\times10^{-3}$ & $\alpha_{4}\times10^{-3}$ & $\alpha_{5}\times10^{-3}$ & $\alpha_{6}\times10^{-3}$ \\
\midrule
Hydro with chiral EoS & 1.20$\pm$ 0.70 & 3.70$\pm$ 0.70 & 8.40$\pm$ 0.80 & 14.10$\pm$ 0.90 & 20.80$\pm$ 1.10\\
\hline
Hydro with hadronic EoS &0.20$\pm$ 0.10  & 1.18$\pm$ 0.32  &3.50$\pm$ 0.78  & 5.20$\pm$ 0.94 & 11.38$\pm$ 0.99\\
\hline
Hydro with MM and MB scattering off & 1.26$\pm$ 0.90  & 5.31$\pm$ 1.40 & 10.70$\pm$ 1.46 & 18.10$\pm$ 1.88 & 27.9$\pm$ 2.46\\
\hline
Hydro with resonance decays off  & 1.06$\pm$ 0.61 & 3.82$\pm$ 0.94 & 9.19$\pm$ 1.02 & 17.23$\pm$ 1.38  & 24.56$\pm$ $1.93$\\
\hline
Hydro with both MM, MB and resonsnce decays off & 1.44$\pm$ 0.90  & 5.38$\pm$ 1.46  & 13.25$\pm$ 1.40 & 20.20$\pm$ 1.61  & 31.50$\pm$ 2.54\\
\bottomrule
\end{tabular}
\end{table}

\section{Summary}
\label{sec3}

From the present investigation of two dimensional scaled factorial moments analysis on $\chi(\eta,\phi)$ spaces, it is found that the data generated with UrQMD transport model or random data do not exhibit any noticeable signature of self similarity or intermittency. On the other hand, the data of hybrid UrQMD-hydro model, which is a mixture of transport and hydrodynamic models, does exhibit intermittency both for hadronic and chiral EoS. The observed power law behavior seen in UrQMD-hydro data with both hadronic and chiral EoS, and not in UrQMD data, confirms that the observed intermittency is not associated with the nature of the medium produced in the heavy-ion collision, but on the mechanism of evolution of the medium produced in such collision. The particle production is found to be more preferential in UrQMD-hydro generated data with chiral EoS than that of hadronic EoS. In our effort to assess the effect of final state re-scattering and resonance decays on the strength of the intermittency, it is found that both hadronic re-scattering and resonance decays only weaken the strength of the intermittency. Thus, the many particle correlations that could be observed with our UrQMD-hydro data could not arise due to later stage binary decays.

\section*{Acknowledgements}
The authors thankfully acknowledge the UrQMD group for developing UrQMD and UrQMD-hydro codes that have been used for generating events for this work. The authors also acknowledge the Department of Science and Technology (DST), Government of India, for providing funds to develop a high-performance computing cluster (HPCC) facility, through the Project No. SR/MF/PS-01/2014-GU, which has been used to generated Monte Carlo (MC) events.




\end{document}